\documentclass[preprintnumbers, twocolumn,reprint, nofootinbib,superscriptaddress]{revtex4}
\usepackage{array}
\usepackage{amsmath}
\usepackage{graphicx}
\usepackage{amssymb}

\begin{document}
\global\long\def\six{\mathbf{6}}
\global\long\def\three{\mathbf{3}}
\global\long\def\met{\not{\!{\rm E}}_{T}}

\preprint{ANL-HEP-PR-10-49, EFI-10-24, IPMU10-0168, NUHEP-TH/10-22}

\title{Color Sextet Vector Bosons and Same-Sign Top Quark Pairs at the LHC}
\author{Hao Zhang}
\affiliation{Enrico Fermi Institute, University of Chicago, 
Chicago, Illinois 60637, U.S.A.}
\affiliation{Department of Physics and State Key Laboratory of Nuclear Physics and Technology, 
Peking University, Beijing 100871, China}

\author{Edmond L. Berger}
\affiliation{High Energy Physics Division, Argonne National Laboratory, 
Argonne, Illinois 60439, U.S.A}

\author{Qing-Hong Cao}
\affiliation{Enrico Fermi Institute, University of Chicago, Chicago, 
Illinois 60637, U.S.A.}
\affiliation{High Energy Physics Division, Argonne National Laboratory, 
Argonne, Illinois 60439, U.S.A}

\author{Chuan-Ren Chen}
\affiliation{Institute for Physics and Mathematics of the Universe, 
University of Tokyo, Chiba 277-8568, Japan}

\author{Gabe Shaughnessy}
\affiliation{High Energy Physics Division, Argonne National Laboratory, 
Argonne, Illinois 60439, U.S.A}
\affiliation{Department of Physics and Astronomy, Northwestern University, 
Evanston, Illinois 60208, U.S.A}

\begin{abstract}
We investigate the production of beyond-the-standard-model color-sextet vector bosons at the 
Large Hadron Collider and their decay into a pair of same-sign top quarks. We demonstrate 
that the energy of the charged lepton from the top quark semi-leptonic decay serves as a good 
measure of the top-quark polarization, which, in turn determines the quantum 
numbers of the boson and distinguishes vector bosons from scalars.  

\end{abstract}

\maketitle

{\bf Introduction -}  The cross section for production of top quarks is relatively 
high at the energies of the Large Hadron Collider (LHC).  While conventional mechanisms 
produce either a single top-quark or a top-antitop pair, it is important to be alert to the 
observation of a pair of same-sign top quarks.  One consequence would be the appearance of 
a pair of same-sign leptons with large transverse momentum.  In a recent paper, 
we explore the potential for discovery of an exotic color-sextet 
scalar in the production of a pair of same-sign top-quarks in early runs of the LHC 
at 7~TeV~\cite{Berger:2010fy}.  The standard model (SM) backgrounds are small.  
We demonstrate that one can measure the scalar mass and the top-quark polarization, 
and confirm the scalar nature of the resonance 
with $1\,{\rm fb}^{-1}$ of integrated luminosity.  Moreover, the 
top-quark polarization distinguishes gauge triplet- and singlet-scalars.   
In the present manuscript, we address color sextet {\em vector} production in 
same-sign top-quark pair-production, and we show that its discovery, and 
the determination of its properties, could also be accomplished with early LHC data.  

Important in our analysis is the recognition that among the products of semi-leptonic top-quark decay, the direction of the charged-lepton is highly correlated with the top-quark spin.  
The top-quark polarization can be measured from the distribution 
in $\cos \theta$~\cite{Mahlon:1995zn}, 
the cosine of helicity angle between the charged-lepton momentum in the top-quark rest frame 
and the top-quark momentum in the center-of-mass frame of the production process 
(i.e. the rest frame of the parent scalar or vector). 
Gauge triplet scalars decay to $t_{L}t_{L}$, and gauge singlet scalars to $t_{R}t_{R}$.  
Here, $t_{L}$ and $t_{R}$ denotes top quarks with left-handed and right-handed polarization.
In either case, both top quarks produce the same angular distribution, 
either $(1-\cos\theta)/2$ ($t_Lt_L$)or $(1+\cos\theta)/2$ ($t_Rt_R$), 
allowing unambiguous identification of the scalar~\cite{Berger:2010fy}. 
A color sextet vector decays into a $t_L t_R$ pair.  
In this case, the observed inclusive angular distribution in the final state would be a flat, 
a sum of the shapes from the $t_L$ and the $t_R$ decays.  
The flat profile would distinguish a vector from a scalar, 
but it also would admit a mechanism that yields unpolarized top quarks.  

In this paper we establish that one can separate the angular distributions corresponding to $t_L$ and $t_R$ from the color sextet vector decay into a $t_L t_R$ pair.  The solution relies of the introduction of asymmetric cuts on the momenta of the leptons from the top-quark decays.      

{\bf The Model -} The most general $SU(3)_{C}\times SU(2)_{L}\times U(1)_{Y}$ effective invariant Lagrangian for color sextet scalars $\Phi$ and vectors $V_{\mu}$ has the form~\cite{Atag:1998xq,Arik:2001bc,Cakir:2005iw} 
\begin{eqnarray}
 &  & \mathcal{L}=\left(g_{1L}\overline{q_{L}^{c}}i\tau_{2}q_{L}+g_{1R}\overline{u_{R}^{c}}d_{R}\right)\Phi_{6,1,1/3}\nonumber \\
 &  & \quad+\,\, g_{1R}^{\prime}\overline{d_{R}^{c}}d_{R}\Phi_{6,1,-2/3}+g_{1R}^{\prime\prime}\overline{u_{R}^{c}}u_{R}\Phi_{6,1,4/3}\nonumber \\
 &  & \quad+\,\, g_{3L}\overline{q_{L}^{c}}i\tau_{2}\tau q_{L}\cdot\Phi_{6,3,1/3}\nonumber \\
 &  & \quad+\,\, g_{2}\overline{q_{L}^{c}}\gamma_{\mu}d_{R}V_{6,2,-1/6}^{\mu}+g_{2}^{\prime}\overline{q_{L}^{c}}\gamma_{\mu}u_{R}V_{6,2,5/6}^{\mu}+h.c.\,,\qquad
 \label{eq:lag}
 \end{eqnarray}
where $q_{L}=(u_{L},d_{L})$ denotes the left-handed quark doublet,
$u_{R}$ and $d_{R}$ are the corresponding right-handed gauge singlet
fields, and $q^{c}\equiv C\bar{q}^{T}$ is the charge conjugated quark
field.   For the sake of simplicity, color and generation
indices are omitted.  The subscripts on the $\Phi$ and $V$ fields denote the
standard model gauge quantum numbers ($SU(3)_{C}$, $SU(2)_{L}$, $U(1)_{Y}$)~%
\footnote{The vector $V_{6,2,-1/6}$ is not considered here as it cannot decay
into a top-quark pair.  Its collider phenomenology will be presented 
elsewhere~\cite{Berger:unknown}. %
} .

We are interested in same-sign top-quark pair production via a sextet vector decay. 
Only the axial-vector part of the coupling contributes
while the pure vector coupling vanishes due to the identical quarks. 
The effective coupling of the color sextet vector to a pair of identical quarks ($qq$) is 
\begin{equation}
\mathcal{L}_{\rm int} = \frac{g}{2} K_{ab}^{M} \bar{q}_{a}\gamma_{\mu}\gamma_5q_{b}^{c}V_{M}^{\mu} + \text{h.c.}~, \label{eq:lagrangian}
\end{equation}
where the $K_{ab}^{M}$ are Clebsch-Gordan coefficients; $a$ and $b$ are the color indices in 
the fundamental representation; $M$ is the color index in the sextet representation;  
and $g$ is the coupling strength. 
Without loss of generality, we concentrate on real and flavor-conserving couplings in this work.

The coupling of the vector to two up-type quarks is largely constrained by the 
measurement of $D^{0}-\bar{D}^{0}$ mixing which is affected by the vector at the tree level. 
The $|\Delta C = 2|$ Hamiltonian induced by $V$ is
\begin{equation}
\mathcal{H}_{\Delta C=2} = 
\frac{2g_{uu}g_{cc}}{m_V^2}\biggl(C_3(\mu)Q_3+C_2(\mu)Q_2\biggr).
\end{equation}
The four-fermion operators $Q_2$ and $Q_3$ are
\begin{eqnarray}
Q_2&=&\left(\bar{u}_{L\alpha}\gamma^\mu c_{L\alpha}\right)
\left(\bar{u}_{R\beta}\gamma_\mu c_{R\beta}\right), \\
Q_3&=&\left(\bar{u}_{L\alpha}c_{R\alpha}\right)
\left(\bar{u}_{R\beta}c_{L\beta}\right), 
\end{eqnarray}
and the Wilson coefficients are 
\[C_2(m_V)=-1,~~ C_3(m_V)=2.\] 
The vector's contribution to $\Delta m_D \equiv \left|m_{D^0}-m_{\bar{D}^0}\right|$ is 
\begin{eqnarray}
\Delta m_D &=& \left| \Re\left(\frac{2\left<\bar{D}^0|\mathcal{H}|D^0\right>}{2m_D}\right) \right|
\nonumber \\
&=& \frac{1}{m_D} \frac{2 g_{uu}g_{cc}}{m_V^2} 
\biggl|C_2 \left<Q_2\right> + C_3 \left<Q_3\right>\biggr|,
\end{eqnarray}
where the hadron matrix elements are~\cite{Golowich:2007ka}
\begin{equation}
\left<Q_2\right>=-\frac{5}{6}f_D^2 m_D^2 B_D,
~\left<Q_3\right>=\frac{7}{12}f_D^2 m_D^2 B_D,
\end{equation} 
with $f_D=222.6\pm 16.7^{+2.3}_{-2.4}~\rm{MeV}$, $m_D=1865~\rm{MeV}$, 
and $B_D=0.82$~\cite{Artuso:2005ym}.
To be consistent with the measured $\Delta m_D$~\cite{Barberio:2008fa},
\begin{equation}
x_D=\frac{\Delta m_D}{\Gamma_D}\sim 8\times 10^{-3},~\Gamma_D=1.6\times 10^{-12}~\rm{GeV},
\nonumber
\end{equation}
the coupling $g_{qq}$ is stringently constrained:
\begin{equation}
g_{uu}g_{cc}\lesssim 1.6\times10^{-8}
\end{equation}
for $m_V \simeq 1~{\rm TeV}$, after an enhancement factor $\sim 2.1$ is included from
the QCD running of the Wilson coefficients from the scale $m_V$ to $m_c$. 

This strong constraint on the product $g_{uu}g_{cc}$ allows freedom for the production
of $V$ at the LHC if the coupling $g_{cc}$ to the second generation quarks is minimized.  
Alternatively, the first generation may be suppressed while the second generation is not, 
but such sea-quark initiated processes are relatively suppressed.

\begin{figure}
\includegraphics[clip,scale=0.6]{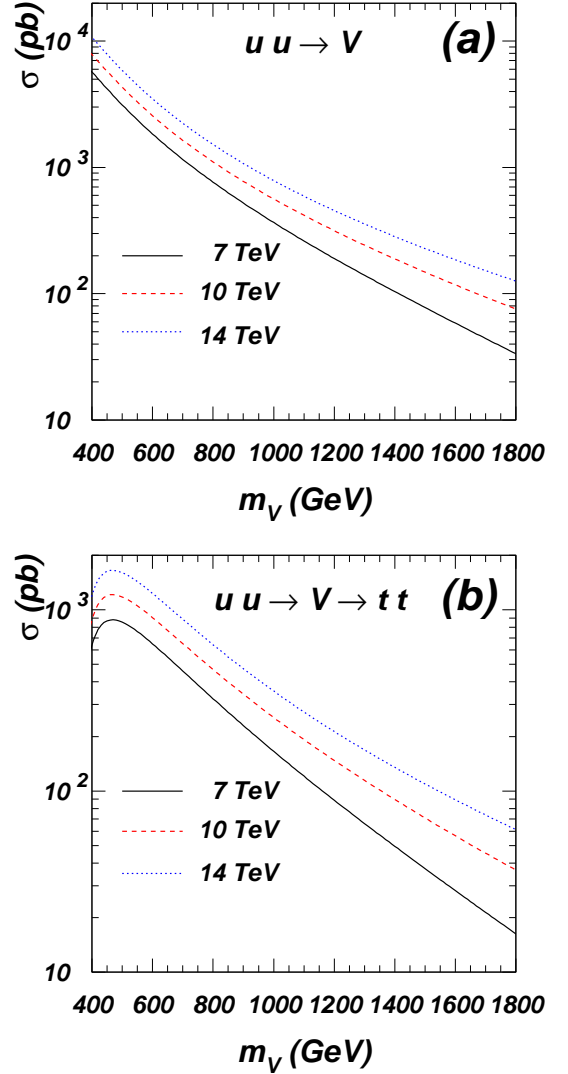}
\caption{Leading order cross sections (pb) for (a) $V$ production via $uu\to V$ 
and (b) like-sign $t$-pair production via $uu \to V \to t t$ at LHC energies: 7~TeV
(solid), 10~TeV (blue dashed), and 14~TeV (red dashed) for $g=1$.
We use the CTEQ6L parton distribution functions~\cite{Pumplin:2002vw}
and choose the renormalization and factorization scales as $m_{V}$.
\label{fig:xsec}}
\end{figure}

For the choice $g=1$, we show the cross section for $V$ production via the 
process $uu\to V$ at the LHC in 
Fig.~\ref{fig:xsec}(a) and the $t t$ cross section via the process $uu\to tt$ in Fig.~\ref{fig:xsec}(b).  
The large cross sections arise from the large parton distribution functions for valence $u$ 
quarks in the initial state.  If the LHC energy is raised from 7~TeV to 14~TeV the cross section 
is increased by roughly a factor of 3 to 4. 

{\bf Discovery Potential -} We focus on 
the same sign dilepton decay mode in which the $W$ bosons from both $t \rightarrow W b$ decays 
lead to a final state containing an electron or muon, $W \rightarrow l \nu$, accounting for 
about $5\%$ of all $tt$ decays.
We concentrate on the clean $\mu^{+}\mu^{+}$ final state because
muon reconstruction has a large average efficiency of $95-99\%$ within the  
pseudorapidity range $\left|\eta\right|<2.4$ and transverse momentum range 
$5\text{ GeV}\le p_{T}\le 1\text{ TeV}$, 
while the charge mis-assigned fraction for muons with $p_{T}=100\,{\rm GeV}$
is less than $0.1\%$~\cite{Aad:2009wy}. 

\begin{table*}
\caption{Signal and background cross sections (pb) before and after cuts, with $g=1$, for six values of $m_{V}$~(GeV).   The decay branching ratios of the 
signal ${\rm Br}(tt)$ are given in the second column.  The ``no cut'' rates correspond to all lepton 
and quark decay modes of $W$-bosons, whereas those  ``with cut''  are obtained after all cuts, the 
restriction to 2 $\mu^+$'s and with tagging efficiencies included. 
\label{tab:xsec}}
\begin{tabular}{cccc|cccc|cccc}
\hline 
$m_{V}$ & ${\rm Br}(tt)$  & No cut  &  With cut &$m_V$ & ${\rm Br}(tt)$  & No cut  &  With cut &Background & No cut &  With cut  \tabularnewline
\hline
500  & 0.446  & 10.97  & 1.71 & 800  & 0.483  &  4.22  & 1.21   & $t\bar{t}$  & 97.62   & 0.0004 \tabularnewline
600  & 0.466  &  8.22  & 1.76 & 900  & 0.487  &  3.02  & 0.92   & $WWjj$      &  9.38   & 0.00001 \tabularnewline
700  & 0.477  &  5.89  & 1.53 & 1000 & 0.489  &  2.22  & 0.70   & $WWW/Z$     &  0.03   & 0.0006 \tabularnewline
\hline
\end{tabular}
\end{table*}

These events are characterized by two high-energy same-sign leptons,
two jets from the hadronization of the $b$-quarks, and large missing
energy ($\not{\!\!{\rm E}}_{T}$) from two unobserved neutrinos. We generate the dominant backgrounds
with ALPGEN~\cite{Mangano:2002ea}:
\begin{eqnarray}
pp&\to& W^{+}(\to \ell^+\nu)W^{+}(\to \ell^+\nu)jj, \\
pp&\to& t\bar{t} \to bW^{+}(\to \ell^+\nu)\bar{b}(\to \ell^+)W^{-}(\to jj).
\end{eqnarray}
The first process ($WWjj$) is the SM irreducible background
while the second ($t\bar{t}$) is a reducible background as it contributes
when some tagged particles escape detection, carrying small
$p_{T}$ or falling out of the detector rapidity coverage.  For example,
one of the $b$-quarks
decays into an isolated charged lepton while one of the two jets from the $W^{-}$
boson decay is mis-tagged as a $b$-jet. 
Other SM backgrounds, e.g. triple gauge boson production 
($WWW$, $ZWW$, and $WZg(\to b\bar{b})$), 
occur at a negligible rate after kinematic cuts, and are not shown here.

At the analysis level, all signal and background events are required to pass the
following acceptance cuts:
\begin{eqnarray}
&&p_T^j\geq 50\,{\rm GeV}, \quad \left|\eta_{j}\right|\leq 2.5 \nonumber \\
&&p_{T}^{\ell_{\rm greater}}\geq 50\,{\rm GeV},\quad p_{T}^{\ell_{\rm lesser}}\geq 20\,{\rm GeV},\quad
\left|\eta_{\ell}\right|\leq2.0, \nonumber \\
&&\Delta R_{jj,j\ell,\ell\ell} > 0.4,
\label{eq:cut}
\end{eqnarray}  
where we order the two charged leptons in the final state by their energies and label the more
energetic lepton as ``greater'' and the other one as ``lesser''.   Owing to spin correlations,
the charged lepton from right-handed top quark decay is more energetic than the one from the 
left-handed top quark decay.  This difference motivates our asymmetric cut on the $p_T$ of the two charged leptons.  The separation $\Delta R$ in the azimuthal angle ($\phi$)-pseudorapidity ($\eta$) plane between the objects $k$ and $l$ is
\begin{equation} 
\Delta R_{kl}\equiv\sqrt{\left(\eta_{k}-\eta_{l}\right)^{2}+\left(\phi_{k}-\phi_{l}\right)^{2}}.
\end{equation}
We model detector resolution effects by smearing the final
state energy according to
\begin{equation} 
\frac{\delta E}{E}= \frac{\mathcal{A}}{\sqrt{E/{\rm GeV}}}\oplus \mathcal{B},
\end{equation}
where we take $\mathcal{A}=10(50)\%$ and $\mathcal{B}=0.7(3)\%$ for leptons(jets). To
account for $b$-jet tagging efficiencies, we demand two $b$-tagged jets, each with
a tagging efficiency of $60\%$.
We also apply a mistagging rate for charm-quarks 
$\epsilon_{c\to b}=10\%$ for $p_{T}(c)>50\,{\rm GeV}$.
The mistag rate for a light jet is
$\epsilon_{u,d,s,g\to b}= 0.67\%$ for $p_{T}(j)<100\,{\rm GeV}$
and $2\% $ for $p_{T}(j)>250\,{\rm GeV}$.
For $100\,{\rm GeV}<p_{T}\left(j\right)<250\,{\rm GeV}$,
we linearly interpolate the fake rates given above.

After lepton and jet reconstruction, we demand that the two hard leptons are of the same sign, 
a requirement which greatly reduces the SM background, 
giving a rejection of order $10^{-4}$ and $10^{-3}$ for 
the $t\bar t$ and $WWjj$ processes, respectively.  
After the cuts are imposed, we find a total of 1.0 background event, 
0.4 from $t\bar t$ and 0.6 from $WWjj$ for 1 fb$^{-1}$ of integrated luminosity.  

After $b$-tagging and restriction to the $\mu^+\mu^+$ mode, 
15-30\% the signal events survive the analysis cuts depending on the vector mass.
Signal and background cross sections are shown 
in Table~\ref{tab:xsec}, before and after cuts, for 6 values of $m_V$.

Because the decay width of $V$ is narrow, 
\begin{equation}
\Gamma(V\to qq) = \frac{g^2 m_V}{24\pi} \left(1-\frac{4 m_q^2}{m_V^2}\right)^{3/2},
\end{equation}
one can factor the process $uu\to tt$ into vector production and decay 
terms,  
\begin{eqnarray}
&& \sigma(uu\to V \to tt) =\sigma_{0}(uu\to V) \times g^2 {\rm Br}(tt), 
\nonumber \\
&=& \sigma_{0}(uu\to V \to tt) \times g^2 \frac{{\rm Br}(tt)}{{\rm Br}_0(tt)}.
        \label{eq:conv}
\end{eqnarray}
The decay branching ratio ${\rm Br}(tt)\equiv{\rm Br}(V \to tt)$ is 
\begin{eqnarray}
{\rm Br}(tt)=\frac{g_{tt}^2 R}{g_{uu}^2+ g_{tt}^2 R}&,& R = (1-4 m_t^2/m_V^2)^{3/2}.
\label{ea:br}
\end{eqnarray}
Subscript ``0" denotes the reference value $g=1$.  
We choose to work with the following two parameters
in the rest of this paper: the vector mass $m_V$ and the product 
$g^2 {\rm Br}(V \to tt)$.  The kinematics of the final state 
particles are determined by the vector mass, whereas the couplings of the 
vector to the light and heavy fermions change the overall normalization.

\begin{figure}[t]
\includegraphics[scale=0.46,clip]{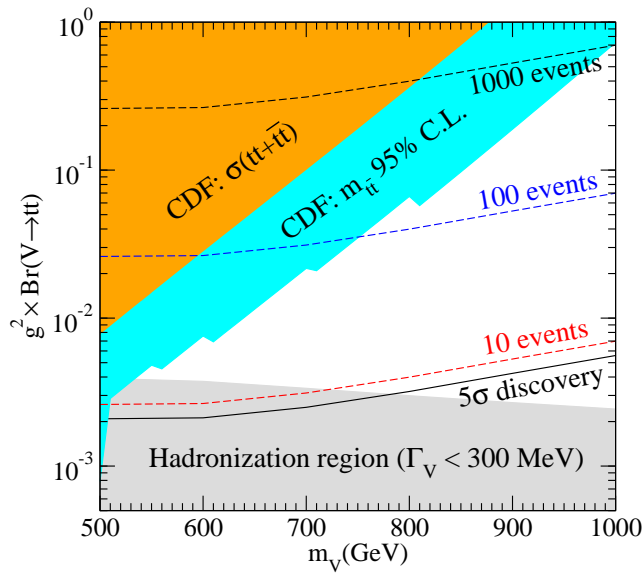}
\caption{Event number contours as a function of the vector mass and the parameter 
$g^2 \text{Br}(V\to tt)$ after all cuts with an integrated luminosity of $1~{\rm fb}^{-1}$.
The shaded regions are excluded, as explained in the text. 
\label{fig:discovery}}
\label{reach}
\end{figure}

In Fig.~\ref{fig:discovery}, we show the expected numbers of signal events as 
a function of $m_V$ for a range of values of the coupling $g^2 \text{Br}(V\to tt)$.  
We obtain the event rate lines by converting the required cross section into 
$g^2 \text{Br}(V\to tt)$ via Eq.~\ref{eq:conv}.  
Based on Poisson statistics, one needs 8 signal events in order to claim a 
$5\sigma$ discovery significance (equivalent to a 99.999943\% confidence level) 
on top of 1 background event. We plot the $5\sigma$ discovery line (solid) in the figure. 
Rates for other values of $g_{uu}$ and $g_{tt}$ can be obtained from Eq.~\ref{eq:conv}.

The search for same-sign top quark pair production in the dilepton 
mode at the Tevatron imposes an upper limit 
$\sigma(tt+\bar{t}\bar{t})\leq 0.7\,{\rm pb}$~\cite{BarShalom:2008fq,Aaltonen:2008hx,Cao:2010zb}.
The constraint is plotted in the orange shaded region.  
The CDF collaboration measured the $t\bar{t}$ invariant mass
spectrum in the semi-leptonic decay mode~\cite{Aaltonen:2009iz}.  Since $b$ and 
$\bar{b}$ jets from $t \rightarrow W b$ are not distinguished well, $tt$ pairs 
lead to the same signature as $t\bar{t}$ in the semi-leptonic mode.  Hence, 
the $m_{t\bar{t}}$ spectrum provides an upper limit on $\sigma(tt+\bar{t}\bar{t})$,
shown in 
the cyan shaded region in Fig.~\ref{fig:discovery}.  The lower gray 
shaded region is the region in which $V$ would 
hadronize before decay, washing out the spin correlation effects we 
utilize to probe the coupling and spin of the sextet state.

{\bf Top Quark Polarization -}~
We use the observation of a pair of same sign dileptons as indicative of a same sign top quark
pair and to suppress SM backgrounds efficiently.   The two missing neutrinos in the final state  complicate event reconstruction.  Following Ref.~\cite{Berger:2010fy}, we use the MT2 method 
to select the correct $b$-$\mu$ combinations and to verify whether the final state is 
consistent with $t \to Wb$ parentage.   Then we make use of the on-shell
conditions of the two $W$ bosons and two top quarks to solve for 
the neutrino momenta~\cite{Sonnenschein:2006ud,Bai:2008sk}.  
Once the neutrino momenta are known, the kinematics of the entire final state is fixed, and the 
vector boson mass is computed from the invariant mass of the two reconstructed top quarks. 

\begin{figure}[t]
\includegraphics[scale=0.43,clip]{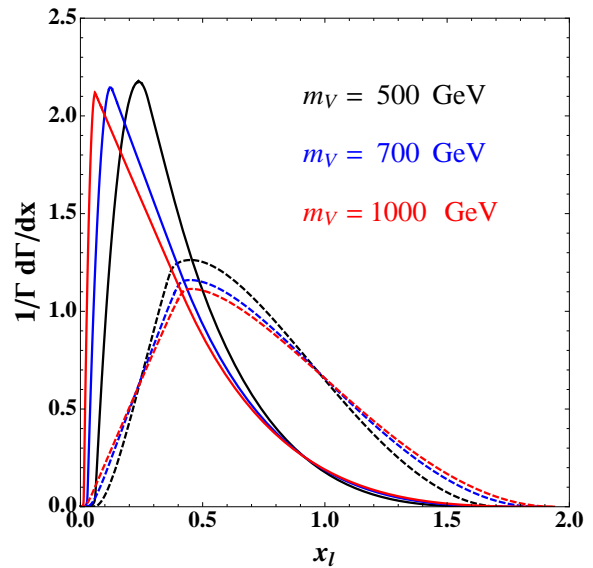}
\caption{Normalized distributions of the energy of the charged lepton from top-quark decay
for $m_V=500,~700,~1000~{\rm GeV}$. 
The solid lines correspond to the left-handed top-quark decay while the dashed lines
to the right-handed top-quark decay. 
\label{fig:en_lep} }
\end{figure}

The next step is to verify that the top-quarks exhibit opposite polarization, accomplished here by 
making use of the difference in the momentum spectra of decay leptons from left-handed and 
right-handed top quarks~\cite{Schmidt:1992et}.  The energy of a charged lepton in a right-handed 
top-quark decay is harder than the one in a left-handed top-quark decay. 
The differential distribution in the energy of the charged lepton energy is
\begin{eqnarray}
\frac{d\Gamma}{dx} &=& \int^{z_{\rm max}}_{z_{\rm min}} dz~ 
\frac{\partial(x^\prime z^\prime)}{\partial(x,z)}~ \frac{d\Gamma}{dx^\prime dz^\prime},
\end{eqnarray}
where $x=2E_\ell/E_t$ is the energy fraction of the charged lepton, and $z=\cos\theta$ where 
$\theta$ is the helicity angle defined in the Introduction. The variables $x^\prime$ and $z^\prime$ are defined in the top-quark rest frame
and are linked to $x$ and $z$ in the laboratory frame through the top quark boost:
\begin{equation}
x^\prime = x \gamma^2 (1-z\beta), \quad z^\prime  = \frac{z-\beta}{1-z \beta},
\end{equation}
where $\gamma=E_t/m_t$ and $\beta=\sqrt{1-1/\gamma^2}$. 
The lower and upper limits of integration are 
\begin{eqnarray}
z_{\rm min}&=&{\rm Max}\left[\frac{1}{\beta}\left(1-\frac{1}{\gamma^2 x}\right),-1\right],\\
z_{\rm max}&=&{\rm Min}\left[\frac{1}{\beta}\left(1-\frac{B}{\gamma^2 x}\right),~1\right].
\end{eqnarray}
The differential cross section $d\Gamma/dx^\prime dz^\prime$ in the rest frame of the top quark 
is~\cite{Jezabek:1988ja}
\begin{eqnarray}
\frac{d\Gamma}{dx^\prime dz^\prime} &=& \frac{\alpha_w^2}{32\pi}
\frac{m_t}{AB} x^\prime (1-x^\prime) \nonumber \\
&\times &
{\rm ArcTan}\left[
\frac{A x^\prime}{B - x^\prime}
\right]
\frac{1+\hat{\mathbf{s}}_t z^\prime}{2},\label{eq:en_topframe}
\end{eqnarray}
where $A=\Gamma_W/m_W$, $B=m_W^2/m_t^2$,
\[
{\rm ArcTan}(x)=
\begin{cases}
\arctan(x), & {\rm for}~x\geq 0,\\
\pi+\arctan(x), & {\rm for}~x<0,
\end{cases}\]
and $\hat{\mathbf{s}}_t$ labels the top quark spin direction.
Here, $m_W$ and $\Gamma_W$ denote the mass and width of the $W$ boson, respectively.

\begin{figure}[b]
\includegraphics[scale=0.6,clip]{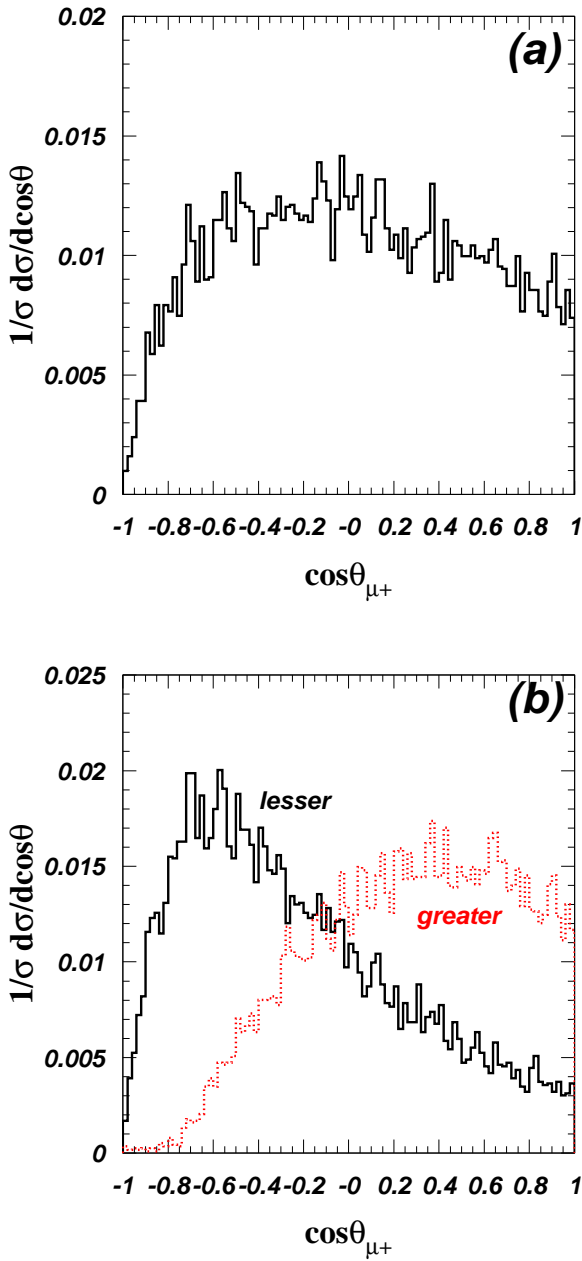}
\caption{(a) Distribution in $\cos\theta$ for reconstructed top quarks 
from $V\to t_R t_L$ without energy selection 
on the decay leptons.   Both $\mu^+$ leptons exhibit similar distributions.  
(b) Distribution in $\cos\theta$ for the reconstructed $t_{{\rm lesser}}$ (black-solid)
and $t_{{\rm greater}}$ (red-dashed). For illustration we choose $m_V = 600~{\rm GeV}$.
\label{fig:pol}}
\end{figure}

In this work we choose the helicity basis to measure the top quark polarization. 
In this basis, the top quark spin is chosen to be along (against) the direction of motion 
of  the top quark in the center of mass frame of the system.   After a boost from the top-quark 
rest frame to the laboratory frame,
we obtain the energy distributions of the charged leptons from left-handed and 
right-handed top-quark decays shown in Fig.~\ref{fig:en_lep}.  The solid curves denote
the $t_L$ decay while the dashed curves the $t_R$ decay.   
As the charged lepton follows the top quark spin, the lepton from $t_R$ decay tends to follow 
the direction of motion of the top-quark, and is more energetic.  The lepton from
the $t_L$ decay tends to move against the direction of motion of its top quark, and it is 
therefore less energetic.  The difference in the energy spectra becomes more evident with 
increasing $m_V$.  For example, for a 1000~GeV vector, the charged lepton from $t_R$ 
decay peaks near $x=0.5$ while the one from $t_L$ decay peaks below $x=0.1$. 
We note that both solid and dashed curves have a kink feature.  
For the left-handed top-quark decay, the kink arises largely from the boost, 
which generates the lower integration limit $z_{\rm min}$. The limit  
yields a scale of $x\simeq 1-\beta$ for a heavy vector.  
On the other hand, the kink in the distribution for  
right-handed top-quark decays comes mainly from the non-continuity of the ${\rm ArcTan}$ function in 
the matrix element, which yields a scale of $x \simeq B~(1+\beta)$ after the boost.

\begin{figure}[b]
\includegraphics[scale=0.55]{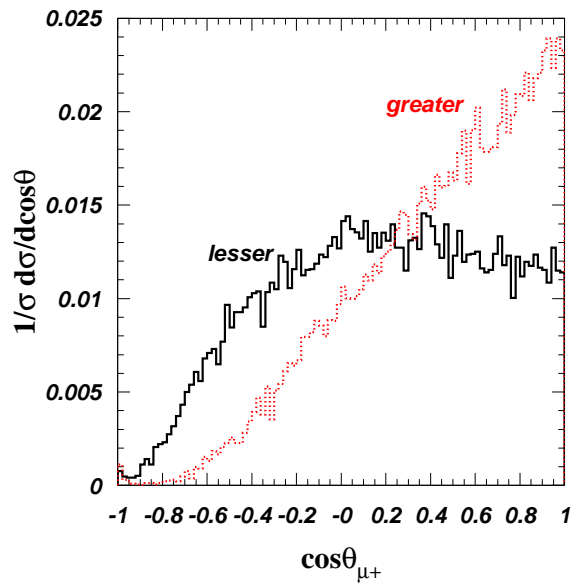}
\caption{Distribution of $\cos\theta$ of the reconstructed $t_{{\rm lesser}}$
and $t_{{\rm greater}}$ of a singlet scalar decay into $t_R t_R$. 
For illustration we choose $m_V = 600~{\rm GeV}$.
\label{fig:pol-scalar}}
\end{figure}

In sextet vector boson decay into a $t_L t_R$ pair,  the charged leptons exhibit a mixture of 
the energetic and soft spectra. To utilize this feature,  we order the energy of the 
leptons and define 
the top quark containing the greater energy lepton as $t_{{\rm greater}}$
and the other top quark as $t_{{\rm lesser}}$.   The $\cos\theta$ distributions
of the reconstructed $t_{{\rm lesser}}$ and $t_{{\rm greater}}$ are displayed
in Fig.~\ref{fig:pol}.   These results show that one can differentiate the 
$(1+\cos\theta)$ ($t_R$) and $(1-\cos\theta)$ ($t_L$) shapes.   The reconstruction 
is not perfect, however, owing to imperfect assignment of the charged lepton.  
As shown in Fig.~\ref{fig:en_lep},  charged leptons from $t_L$ decay can be 
more energetic than those from  $t_R$ decay with small probability.  With 
increasing vector mass,  the probability of wrong assignment becomes smaller.  

As a consistency check on this method for determining polarizations, we apply 
the same reconstruction to singlet scalar decay into  two right-handed top-quarks. 
The reconstructed $t_{{\rm lesser}}$ and $t_{{\rm greater}}$ distributions for  
singlet scalar decay are shown in Fig.~\ref{fig:pol-scalar}.  Both distributions 
maintain the expected $(1+\cos\theta)$ trend, but the shapes are distorted by 
event reconstruction.   Hence, we gain added confidence that the top-quark 
polarization can be determined and used to discriminate vector bosons from 
scalar bosons.  

{\bf Summary -} We present a study of the search for exotic charge $4/3$ 
color-sextet vector bosons in the production of same-sign top-quark pairs at the 
LHC at 7~TeV.   We examine the final states in which both top-quarks decay 
semi-leptonically.  We show that vector bosons can be distinguished from 
scalars.  The top quarks from vector decay exhibit opposite polarization, one 
left-handed and one right-handed.  The inclusive distribution in the helicity 
angle $\cos\theta$ of a charged lepton is flat since it receives contributions 
from both top quarks.  However, we show that an energy selection on the 
charged leptons can restore the characteristic shapes that distinguish 
left- and right-handed top quarks.  Correspondingly, we show that LHC 
data should allow one to observe $V \to t_R t_L$ and distinguish vector 
from scalar decay.  

{\bf Acknowledgments -} The work by E.L.B., Q.H.C. and G.S. is supported in part by the U.S.
Department of Energy under Grant No.~DE-AC02-06CH11357. Q.H.C is also
supported in part by the Argonne National Laboratory and University
of Chicago Joint Theory Institute Grant 03921-07-137. C.R.C. is supported
by World Premier International Initiative, MEXT, Japan. G.S. is also
supported in part by the U.S. Department of Energy under Grant No. DE-FG02-91ER40684.
H.Z. is supported in part by the U.S. Department of Energy under Grant 
No.~DE-FG02-90ER40560 and also in part by the National Natural Science Foundation
of China under Grant 10975004 and the China Scholarship Council File No. 2009601282.
Q.H.C thanks Shanghai Jiaotong University for hospitality where part of this work
was done.  
 
\bibliography{reference}

\end{document}